\documentstyle[aps,epsf,prl,multicol]{revtex}
\begin{document}

\newcommand{\be}{\begin{equation}}
\newcommand{\ee}{\end{equation}}
\newcommand{\ba}{\begin{eqnarray}}
\newcommand{\ea}{\end{eqnarray}}
\newcommand{\bt}{\begin{tabular}}
\newcommand{\et}{\end{tabular}}
\newcommand{\bc}{\begin{center}}
\newcommand{\ec}{\end{center}}
\newcommand{\ben}{\begin{enumerate}}
\newcommand{\een}{\end{enumerate}}
\newcommand{\bi}{\begin{itemize}}
\newcommand{\ei}{\end{itemize}}
\newcommand{\bmpage}{\begin{minipage}}
\newcommand{\empage}{\end{minipage}}
\newcommand{\disty}{\displaystyle}

\def \al{\alpha}
\def \df {{\rm d}}
\def \del{\delta}
\def \dg{\dagger}
\def\eps{\epsilon}
\def\veps{\varepsilon}
\def\ga{\gamma}
\def \si {\sigma}
\def\la {\lambda}
\def\latl{\tilde{\lambda}}
\def\La {\Lambda}
\def\om {\omega}
\def\Om {\Omega}
\def\K {\tilde{K}}
\def\J {\tilde{J}}
\def\kgr{\bf k}
\def\Pgr{\bf P}
\def\Qgr{\bf Q}
\def\Rgr{\bf R}
\def\Zgr{ {\bf Z} }
\def\bul{$\bullet\;\;$}
\def\Ccal{ {\cal C}}
\def\Hil{ {\cal H} }
\def\Bcal{ {\cal B}}
\def\Scal{ {\cal S}}
\def\Pcal{ {\cal P}}

\def\ad{{\rm ad} }
\def\tef{t_{\rm ef}}

\title{Ground States of the Falicov-Kimball model\\
        with correlated hopping}

\author{J. Wojtkiewicz$^{1}$ and R. Lema\'nski$^{2}$\\[2ex]
\em $^1$Department for Mathematical Methods in Physics,\\
\em Warsaw University, Ho\.za 74, 00-682 Warszawa, Poland,\\
\em $^2$Institute of Low Temperatures and Structure Research,\\
\em Polish Academy of Sciences, P.O.Box 1410, 50-950 Wroc\l aw, Poland}
\maketitle

\begin{abstract}
Two-dimensional spinless Falicov-Kimball model (FKM) with correlated
hopping is studied perturbatively in the limit of large on-site Coulomb
interaction $U$. In the neutral case the effective Hamiltonian in spin variables
is derived up to terms proportional to $U^{-3}$. Unlike the simplest
FKM case, it contains odd parity terms (resulting from the correlated hopping)
in addition to even parity ones. The ground-state phase diagram
of the effective Hamiltonian is examined in the $(a/t, h)$ plane,
where $a/t$ is a parameter characterizing strength of the correlated hopping and
$h$ is a difference of chemical potentials of two sorts of particles present
in the system. It appears to be asymmetric with respect to the change $h\to -h$
and a new ordered phase is found for a certain interval of $a/t$.
\end{abstract}

\begin{multicols}{2} \narrowtext
{\bf Introduction.}
One of the most fascinating, but still mysterious phenomenon observed
in some materials is a charge ordering \cite{THCMCC}. Despite of its oddity,
originated from the quantum nature of interacting electrons, it seems to be
very likely that an inhomogeneous charge distribution is very common effect present
in strongly correlated electron systems  \cite {GROP}. Then, presumably,
it must be related to a number of various phenomena found in the systems
(such as {\em  metal-insulator phase transition, high $T_c$  supecoductivity,
giant magnetoresistivity}, just to mention a few). The above arguments point out
how much important is understanding of a nature of the effect and justify growing
interest in its theoretical description \cite {GROP}. In particular
it is important to determine factors deciding about what sorts of charge
superstructures are formed.

One of the simplest models suitable to decribe charge ordered phases
on a microscopic level is the Falicov-Kimball model, previously applied
to account for the metal-insulator transition \cite{FK}, mixed valence
phenomena \cite{Khom}, crystallization and alloy formation \cite{KennLieb} etc.
Indeed, it was shown that ground state phase diagrams of the simplest version of the
FKM have extremely rich structure \cite{GUJ,LLJGJL,WL}. The great advantage 
of the model is that it is amenable to rigorous analysis\cite{GM,Kennedy}.

However, the simplest version of the FKM, although non-trivial, is not able 
to account for all aspects of real experiments. This is way an objective 
of our studies is gradual inclusion of those terms that were ignored 
in the simplest version of the FKM, yet keeping the model tractable.

In this contribution we investigate the FKM with so-called {\em correlated
hopping term} added (chFKM). This term was already mentioned by Hubbard 
in \cite{Hubbard}. More than a decade ago Hirsch pointed out that
the term may be relevant in explanation of superconducting
properties of strongly correlated electron systems
(he named it {\em the  bond-charge} interaction) \cite{Hirsch89}.
For a last few years some other authors
examined a role of the correlated hopping in the FKM
\cite{RefchFKM}, and the Hubbard model \cite{RefchHM,SV},
mainly in a context of the metal-insulator phase transition.
Inclusion of this term makes electron hopping rate dependent on
occupation numbers of those sites between which an electron hops.

One of the most difficult problems that one encounters, when trying to describe
correlated electron systems, is a choice of reliable method, that enables
to treat the model under consideration in a controllable way. Here we use
a perturbative method valid in the large $U$ limit,
that permits to transform an initial Hamiltonian, having
a small quantum part, into an effective classical one.
The method has been reported in a series of papers by Datta, Fernandez,
Fr{\"o}hlich and Rey-Bellet \cite{DFF1,DFF2,DFF3,DFF4}.
One can use this method to generate a perturbative series up to an arbitrarily
high order in $1/U$, to establish the {\em convergence} of the whole procedure, 
and -- in some cases -- to obtain phase
diagrams in low (but nonzero) temperatures. This can be done by extending
the techniques of Pirogov-Sinai theory \cite{PS} to quantum models.

The aim of our paper is to examine properties of the chFKM in the
{\em perturbative regime}, i.e. in the range of parameters
where all kinds of hopping terms are small in comparison with the on-site
Coulomb interaction term $U$. We are particularly interested in examination of
how correlated hopping term influences charge ordering.
In the first part of our work we derived
an effective Hamiltonian, that is legitimated in the large $U$
limit. Then we found its ground state properties by means of the method of
restricted phase diagrams, used previously to the simplest version of the FKM
\cite{LLJGJL,WL}.

{\bf The model.}
We are dealing with two types of particles defined on a $d$-dimensional simple
cubic lattice $\Zgr^d$: immobile ions and itinerant, spinless electrons.
(Other interpretations of the model have also been considered
\cite{KennLieb,LLJGJL,WL,GM}).

The Hamiltonian defined on a finite subset
$\La$ of $\Zgr^d$ has the form
\be
H_\La = H_{0,\La} + V_\La,
\label{hamcl0}
\ee
where
\be
\disty
H_{0,\La}=    U \sum_{x\in\La} w_{x}n_{x} - \mu_i \sum_{x\in\La} w_{x}
- \mu_e \sum_{x\in\La}n_{x},
\label{hamcl}
\ee
\be
\disty
V_\La= - \sum_{<xy>}[t + a( w_{x}+ w_{y})](c^\dg_{x}c_{y}
+c^\dg_{y}c_x )
\label{hamq}
\ee
Here $c^\dg_{x}$ and  $c_{x}$ are creation and annihilation
operators of an electron at lattice site $x\in\La$, satisfying
ordinary  anticommutation relations and the corresponding
number particle operator is $n_{x} = c^\dg_{x}c_{x}$.
$w_x$ is a classical variable taking values $0$ or $1$.
It measures the number of ions at lattice site $x$.
The chemical potentials of the ions and electrons are $\mu_i$ and $\mu_e$,
respectively, $t$ is the electron hopping amplitude between empty sites
and $a$ is the correlated hopping constant. 
The symbol $<xy>$ denotes an orderless pair of nearest
neighbour sites of the lattice.

In this paper we examine the model in the range of parameters
$t,a << U$. The value of $a$ is usually smaller than that of $t$,
however both these quantities are of the same order. Indeed, in systems
described by the Hubbard-like models, it has been found that
$\mid  a\slash t\mid \approx 0.3$ \cite{Hubbard,Hirsch89}. In our studies
we impose the following condition: $-t\leq a\leq t$ (for $a=0$,
this model reduces to the ordinary FKM).

{\bf General outline of the perturbation scheme.}
The perturbative scheme we use here can be applied to a general class of (lattice)
Hamiltonians (defined on $\La\subset \Zgr^d$) of the following form
\be
H_\La(t)=H_{0,\La} + r V_\La,
\label{hamgen}
\ee
where the unperturbed Hamiltonian $H_{0,\La}$ is a {\em classical} operator
(i.e. it is diagonal in a basis being the product of bases on all
lattice sites) with degenerate ground states. (Obviously the chFKM, introduced
by the formulas above, belong to this class). Our purpose is to
examine the effect of a quantum perturbation  $r V_\La$, where
$r$ is a small parameter. In other words, we want to (block)
diagonalize the Hamiltonian and find its ground states.

To accomplish this task we are looking for an unitary transformation
 $U(r)$,  which (block) diagonalizes the full Hamiltonian.
 In most cases, finding out such a transformation exactly is a hopeless
 job. More constructive method is a {\em perturbative} treatment,
 consisted in ``killing'' the off-diagonal part of perturbation up to some
 finite power of the parameter $r$:
\[
H(r)\equiv H_0+ r V \to \tilde{H}^{(n)}_0(r) + r^{n+1} \tilde{V}^{od},
\]
where $\tilde{H}^{(n)}_0(r)$ is block-diagonal up to the order $n+1$ in $r$.
Following this way one can determine an explicit formula for a diagonal part
of the perturbed Hamiltonian
$\tilde{H}^{(n)}_0(r)$, called the effective Hamiltonian.

Results based on various perturbative schemes have been previously obtained
for the simplest versions (i.e. without correlated hopping term) of the FKM
and Hubbard models: \cite{Kennedy,DFF4,GruberMMU}.
The method we applied in our studies was developed in
\cite{DFF1,DFF2,DFF3,DFF4} for Hamiltonians of the form (\ref{hamgen}).
Their components can have quite general form;
it is sufficient that both of them are sums of finite-range operators
(or even infinite-range but exponentially decreasing with
distance). Moreover, it is assumed that $H_0$ is expressible by
translationally invariant m-potential \cite{DFF2,DFF3,Slawny}.

The technique developed in \cite{DFF1,DFF2,DFF3,DFF4}
has also two other important aspects. First, since we restrict ourselves
to the low-temperature region of the phase diagram, we need to diagonalize only
 {\em a low-energy} part of the Hamiltonian, what considerably simplifies
 the calculations. Second, special care is taken to the form of the transformed
Hamiltonian: it is formulated as a sum of {\em local} operators.
It is necessary to obtain uniform estimates  (i.e. independent of volume) and, 
as a consequence, to establish the convergence of the whole procedure, 
and (in some cases) to examine orderings emerging in the system.

All  this formalism, its background, achievements and limitations can be found in 
\cite{DFF1,DFF2,DFF3,DFF4}. It must be stressed that many analogous results 
have been also obtained by Kotecky and coworkers \cite{Kotecky}.

{\bf Effective Hamiltonian.}
The Hilbert space of the whole system $\Hil_\La$ is a tensor product:
$\Hil_\La = \disty \mathop{\bigotimes}_{x\in\La} \Hil_x$. Every
$\Hil_x$ is spanned by the  states: $|w_x,n_x\rangle$. There are four base
vectors:$|0,0\rangle$,
$|1,0\rangle$, $|0,1\rangle$ and  $|1,1\rangle$.
The corresponding energies are: $0; - \mu_i; - \mu_e; U-\mu_i - \mu_e$.

Let us begin our analysis from the classical part of the
Hamiltonian. It is identical to a classical part of the Hubbard model
and the FKM, and it is well known \cite{DFF4}. The phase diagram 
consist of four regions.
In region $I$, defined by $\mu_i<0,\mu_e<0$, all sites are empty. In
two twin
regions $II_i, II_e$ given by conditions:
$II_i$:
 $\mu_{i}>0$, $\mu_{i} > \mu_{e}$, $\mu_{e}<U$ (for $II_e$, one should
interchange the subscripts $i$ and $e$)
all  sites are in the  $|1,0\rangle$
(corresp. $|0,1\rangle$) state. In the region $III$, given by:
$\mu_i>U, \mu_e>U$, all sites are doubly occupied.
 The most interesting situation is in the neighbourhood of the
 $\mu_i = \mu_e$ line between regions $II_i$ and $II_e$,
 which corresponds to the half-filled band,
where there is a macroscopic degeneracy. We will analyse mainly this
region.

After some relatively straightforward but lengthy calculations,
performed partially with an aid of symbolic computation programs, we have
obtained for $d=2$ the following effective Hamiltonian up to the second order 
of the perturbation theory, i.e. up to terms proportional to $U^{-3}$:
\newpage
\[
H^{(2)}_{\La, \rm eff} = \disty
(h-20\frac{a t^3_{\rm ef}}{U^3}) \sum_{i}s_i
+ \left(2\frac{ t^2_{\rm ef}}{U}
- 18 \frac{ t^4_{\rm ef}}{U^3}
  \right) \sum_{d(i,j)=1}s_i s_j
\]
\[
+
\frac{6 t^4_{\rm ef}+ 8 a^2 t^2_{\rm ef}   }{U^3}
 \sum_{d(i,j)=\sqrt{2}}s_i s_j
+
\frac{4t^4_{\rm ef}+ 2 a^2 t^2_{\rm ef}   }{U^3}
 \sum_{d(i,j)=2}s_i s_j
\]
\[
+
\frac{8 a t^3_{\rm ef}  }{U^3}
  \sum_{\Scal_{3,ijk}}s_i s_j s_k
+
\frac{ 16 a t^3_{\rm ef}  }{U^3}
 \sum_{\Bcal_{3,ijk}}s_i s_j s_k
\]
\be
+
\frac{ 40 t^4_{\rm ef}  }{U^3}
\sum_{\Pcal_{4,ijkl}}s_i s_j s_k s_l
+\frac{3 t^4_{\rm ef} - 10 a^2 t^2_{\rm ef} }{2 U^3}\sum_{\Pcal_{4, ijkl}}
 \bf 1
\label{hamef}
\ee
where: $s_i$ -- the classical one-half spin on the lattice site $i$;
it is related to the variable $w_i$ by the formula:
 $s_i=(w_i-1)\slash{ 2}$;
$t_{\rm ef} = t+a$; $\Bcal_{3,ijk}$ -- ``bent'' triples of spins
$i,j,k$ (i.e. the angle between bonds $ij$ and $jk$ is $\pi\slash~2$);
$\Scal_{3,ijk}$ -- ``straight'' triples; $\Pcal_{4,ijkl}$ is
a  $2\times 2$
plaquette on the lattice; $h =\mu_i - \mu_e$.

{\em Remark 1.} In a general case, where some configurations of the
system are not necessarily half-filled, there are also present  {\em
projections} onto half-filled states. For simplicity, we omit this
aspect here. An expression for the general case we plan to present in
an extended version of this paper.

{\em Remark 2.} For $a=0$, we should obtain an effective
Hamiltonian for the ordinary FKM. Comparing (\ref{hamef}) for
$a=0$  with analogous expression in \cite{DFF4}, Table 2, we
observed full consistence with exception of the constant term (we claim
that authors have omitted  term $\frac{t^4}{2U^3}P^0_{\{ xyzw \}}$
(in their terminology)). But this term is important only for absolute values of
energy; it neither affect the differences of energies, nor the phase diagram.

{\em Remark 3.} The effective Hamiltonian (\ref{hamef}) written in the spin
variables corresponds to the Ising-like model with (dominating) antiferromagnetic
interactions. However there is an important new aspect, as compare with
the simplest FKM, that comes from the correlated hopping term
(i.e. for $a \neq 0$). This is a presence of terms with odd numbers
of the spin operators. As a result the symmetry $h\to -h$
(present in FK and Hubbard models) does not longer  hold.
An additional term proportional to the sum of spin
variables plays a role of a suplementary external field. The other odd terms
can be regarded as generalized fields.

A physical explanation of the reason why these new terms with odd numbers 
of spin operators emerge is straightforward. 
Let us first focus on the linear term. Since in the chFKM the hopping rate
depends on sites occupations, it is energetically favorable if all
occupation numbers have one, out of two possible values: 1 if $a > 0$
(as hopping amplitude between two occupied sites is equal to $t+a$)
and 0 if $a < 0$. (Note that for the simplest FKM $a = 0$ so
the linear term does not contribute to the Hamiltonian, as it is proportional
to $a$). The odd terms of higher order enter in a more subtle way, as they
involve three or more neighbouring sites, but the principal rule stay the same:
it favours an occupation number equal to 1 or 0, for $a > 0$ or
$a < 0$, respectively.

{\bf The phase diagram.}
Since there is no general method of finding ground-states for
classical Hamiltonians (as far as we know), we looked for the ground states
of the Hamiltonian (\ref{hamef}) by minimizing energy in some set of ``trial''
configurations (the method of {\em restricted phase diagrams},
\cite{LLJGJL,WL}).
 We took a set of all {\em periodic} configurations,
having elementary cells up to 12 sites (there are 2000 such nonequivalent
configurations), however it appeared that {\em all} configurations that emerged
in the phase diagram, have no more than 5 sites per elementary cell.
This  observation led us to the conjection, that within the assumed perturbation
order our results are exact, i.e. we claim that other configurations are absent.
We hope that in a future we will be able to find a rigorous proof that our
configurations are true minimizers.

The phase diagram in variables $(a/t,h)$ is displayed in Fig. 1.
It can be noticed that for a predominant set of model parameters the
sequence of phases agrees with that one found for the ordinary FKM
(although the values of $h$ separating subsequent phases depend strongly
on $a$).
However, it is remarkable that for $h<0$ and  $a\in]a_-,a_+[$, where
$a_+ = (-4 - \sqrt{2})\slash 7 \approx -0.773459$, $a_- = (-4 +
\sqrt{2})\slash 7 \approx -0.369398$, a new type of ordering
(labelled by (4)) appears, instead of the three phases (3,5,6).
Consequently, the diagram is clearly asymmetric with respect to the horizontal
axis $h = 0$.

\begin{figure}
\epsfxsize=8.5cm \epsfbox{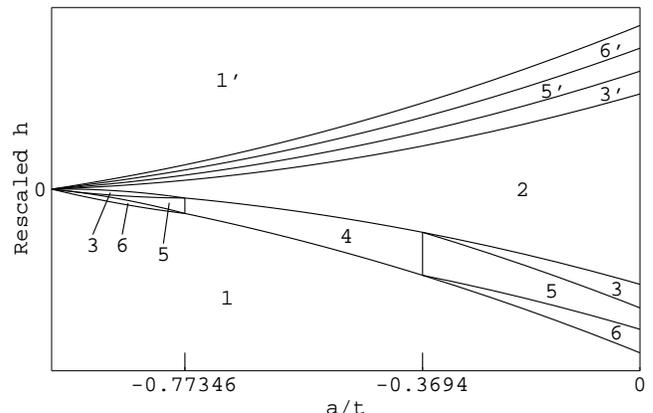}
\caption{\it Schematic phase diagram of the effective Hamiltonian
(\ref{hamef}) of the correlated hopping FKM. Phases represented by various
arrangements of the ions are depicted in Fig. 2.}
\label{fig1}
\end{figure}

Another interesting feature of the diagram is presence of the junction
point for $h = 0,a = -t$, where all lines
separating various phases for $h > 0$, as well as most of the lines for
$h <0$, join together. It worthwhile to notice that the exact solution has
been given just for this characteristic symmetry point \cite{SV}.

\begin{figure}
\epsfxsize=8.5cm \epsfbox{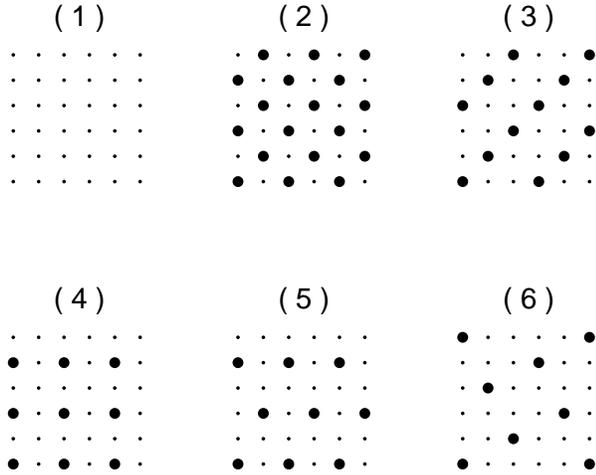}
\caption{\it Configurations of the ions (marked by heavy dots $\bullet$)
corresponding to phases displayed in Fig. 1. Phases labelled by numbers
with prime (e.g. 1', 3', 4' etc.) have mirror configurations with respect to those
without prim, i.e. lattice sites occupied by the ions are then interchanged
with those of unoccupied by the ions.}
\label{fig2}
\end{figure}

{\bf Summary.}
The results obtained in this paper one can summarize as follows. First,
the effective Hamiltonian of the chFKM in the second order perturbation
theory (i.e. up to terms proportional to $U^{-3}$) has been found, and then its
ground state phase diagram has been constructed. It has become evident from our
studies that the correlated hopping term modifies substantionally the effective
Hamiltonian and consequently the phase diagram of the simplest FKM.
The main new feature of the effective Hamiltonian is presence of odd parity
terms. As the result the phase diagram becomes asymmetric with respect to change
of a sign of $h$. In particular, a new ordered phase, that does not exist for
the simplest FKM, has been found for a certain interval of the parameter $a $.

The possible directions for further studies that emerge from our results
are: taking into account subsequent terms of perturbation
theory, investigation the system  at low, but nonzero temperatures and
inclusion additional small terms to quantum part of the Hamiltonian
(for instance, we will allow  hopping of the ions with a small
amplitude $t_i << t_e$, thus obtaining strongly asymmetric Hubbard
model with correlated hopping).

\noindent {\bf Acknowledgements.}

\noindent We acknowledge support from the Polish Research Committee (KBN)
under the Grant No. 2 P03B 131 19.

\end{multicols}

\end{document}